\documentclass[journal]{IEEEtran}
\IEEEoverridecommandlockouts
\usepackage{cite}
\usepackage{balance}
\usepackage{amsmath,amssymb,amsfonts}
\usepackage{algorithm}
\usepackage{algpseudocode}
\usepackage{graphicx}
\usepackage{textcomp}
\usepackage{xcolor}
\usepackage{booktabs}
\usepackage{subfigure}
\usepackage{multirow}
\usepackage{makecell}
\usepackage{xcolor}
\usepackage{soul}
\sethlcolor{white}
\usepackage{tablefootnote}
\usepackage[utf8]{inputenc}
\pdfoutput=1

\begin{document}
\markboth{Journal of Biomedical and Health Informatics}%
{Shell \MakeLowercase{\textit{et al.}}: Bare Demo of IEEEtran.cls for IEEE Journals}

\title{Heterogeneous Collaborative Learning for Personalized Healthcare Analytics via Messenger Distillation}

\author{Guanhua~Ye, Tong~Chen, Yawen~Li, Lizhen~Cui, Quoc~Viet~Hung~Nguyen and  Hongzhi~Yin*
        
\thanks{
G. Ye, H. Yin, and T. Chen are with the School of Information Technology \& Electric Engineering, The University of Queensland, Australia. E-mail: g.ye@uq.edu.au, h.yin1@uq.edu.au, tong.chen@uq.edu.au

Y. Li is with the School of Economics and Management, Beijing University of Posts and Telecommunications, China. E-mail: warmly0716@bupt.edu.cn

L. Cui is with the School of Software, Shandong University, China. E-mail: clz@sdu.edu.cn

Q.V.H. Nguyen is with the School of Information and Communication Technology, Griffith University, Australia. E-mail: henry.nguyen@griffith.edu.au
}
\thanks{* H. Yin is the corresponding author.}
}%

\maketitle

\begin{abstract}
The Healthcare Internet-of-Things (IoT) framework aims to provide personalized medical services with edge devices. Due to the inevitable data sparsity on an individual device, cross-device collaboration is introduced to enhance the power of distributed artificial intelligence. Conventional collaborative learning protocols (e.g., sharing model parameters or gradients) strictly require the homogeneity of all participant models. However,  real-life end devices have various hardware configurations (e.g., compute resources), leading to heterogeneous on-device models with different architectures.  Moreover,  clients (i.e., end devices) may participate in the collaborative learning process at different times.  
In this paper, we propose a \textbf{S}imilarity-\textbf{Q}uality-based \textbf{M}essenger \textbf{D}istillation (SQMD) framework for heterogeneous asynchronous on-device healthcare analytics. By introducing a preloaded reference dataset, SQMD enables all participant devices to distill knowledge from peers via messengers (i.e., the soft labels of the reference dataset generated by clients) without assuming the same model architecture. Furthermore, the messengers also carry important auxiliary information to calculate the similarity between clients and evaluate the quality of each client model, based on which the central server creates and maintains a dynamic collaboration graph (communication graph) to improve the personalization and reliability of SQMD  under asynchronous conditions. Extensive experiments on three real-life datasets show that SQMD achieves superior performance.

\end{abstract}
 

\IEEEpeerreviewmaketitle

\section{Introduction} \label{sec:intro}
\IEEEPARstart{T}{he} evolution of Internet-of-Things (IoT) and Artificial Intelligence (AI) has greatly improved the popularity of many on-device healthcare applications such as sports analytics \cite{gowda2017bringing}, fall detection \cite{chen2006wearable}, and chronic disease monitoring \cite{allet2010wearable}. Among these researches, machine learning \cite{jordan2015machine} (ML) models, especially deep neural networks (DNNs), have shown promising performance when accessing an enormous amount of personal data\cite{chen2018tada}. However, it is usually unrealistic to train a deep predictive model based on a single user's data \cite{zhang2021survey}. The main reason is that the samples collected from one individual are sparse and biased, leading to limited predictive power and over-fitting \cite{chen2020sequence}.

To address the above issue, a common practice is to collect as many users' data as possible on the cloud and train a global model \cite{baker2017internet}. In this centralized setting, the user devices are only used for transmitting locally collected data to the server and displaying cloud-generated results. Consequently, the only way to obtain analytic results is to send a request to the cloud server and wait for the response from the global model, which can be devastating when monitoring fatal diseases under network delay or connection interruptions \cite{ye2022personalized}. Additionally, since the central server collects and stores all personal data, the security of sensitive information and the reliability of the learned models are highly vulnerable to adversarial attacks (e.g., attribute inference attacks \cite{mosallanezhad2019deep} and false data injection \cite{koh2017understanding}). In this regard, on-device distributed learning \cite{imteaj2021survey, zhang2022pipattack, chen2021learning, wang2022fast} was proposed to remove these bottlenecks of the classic centralized paradigm, where the core idea is retaining both the private data and the model training process on devices \cite{vanhaesebrouck2017decentralized}.

Federated Learning (FL) \cite{yang2019federated} has been a performant and representative distributed learning paradigm, given its ability to allow collaborations among clients (e.g., user devices) via a central server. In a general FL setting, each device possesses a private dataset and trains a local model (e.g., DNNs) via corresponding optimization strategies (e.g., stochastic gradient descent) \cite{li2020federated}. Then each client will upload the learned model weights or gradients to the central server. A typical approach of collaborative learning is to aggregate the parameters received from all individual participants (e.g., by taking the average) and then transmit the updated values back to all devices \cite{mcmahan2017communication}. In this way, FL ensures that each device maintains a local model for rapid analysis while learning from all peers without access to their private data.
\cite{reddi2021adaptive} and \cite{lu2022personalized} \hl{further proposed adaptive optimization in FL for scenarios where the data in different clients is imbalanced or heterogeneous.}
However, the conventional FL paradigm requires all participants/clients to share the same model architecture so that the parameters or the gradients from each client can be shared and aggregated on the central server. Real-life IoT devices have various hardware configurations (e.g., compute resources) due to many factors such as different generations of devices or diverse manufacturers, which inevitably leads to the heterogeneity of model architectures on heterogeneous devices. This marks down the practicality of the homogeneous collaborative learning paradigms.

To address the limitation, a few researchers have turned their attention to the heterogeneous FL recently \cite{li2019fedmd}. In the new paradigm, instead of sharing the sensitive data or parameters, clients share their logits (i.e., predicted label probabilities) \hl{w.r.t.} a preloaded reference dataset. The clients then update their models based on both the loss of local training data and the loss of reference data. Sound evidence indicates that such mitigation does transfer knowledge \cite{zhang2021adversarial}. In this way, clients with different model architectures can learn from each other, which is impossible in gradient-based collaborative learning frameworks \cite{geiping2020inverting,wang2020next,zhu2019deep}. The server no longer maintains a global model, instead, it maintains a logits repository. However, the new paradigm encounters the following limitations. 

First, unlike the conventional FL, this heterogeneous FL paradigm requires that all clients participate in the collaborative learning process at the same time. If a new client joins the learning process midway, it cannot immediately catch up with existing peers due to the absence of a global model in this new paradigm, and its poor prediction performance will be propagated across all the clients to hurt the performance of other well-trained client models.  Second,  all clients receive the same knowledge from the central logits repository. That is to say, all clients are fully connected via knowledge distillation so that they converge towards the same direction - the average global logits (i.e., ``single thought''). The resultant models inevitably favor the majority classes of clients and perform poorly for the minority groups of clients.  However, when a minority group of clients reflects marginalized populations like, say, patients over 65 or
Aboriginals, in a blood sugar monitoring app, that bias can become dangerous or even discriminatory.  Despite the efforts to introduce personalized model aggregation into conventional FL \cite{he2018cola, koloskova2019decentralized, zhang2021democratic}, these methods are not applicable to the heterogeneous FL paradigm.
 
To address these two issues, we design a \textbf{S}imilarity-\textbf{Q}uality-based \textbf{M}essenger \textbf{D}istillation (SQMD) framework for heterogeneous asynchronous on-device collaborative learning. 
The key idea of SQMD is that we exploit the important auxiliary information of the logits in addition to the model's dark knowledge. So they are referred to as `messengers' in the paper.
One kind of auxiliary information is the quality of client models. The central server holds the ground-truth labels of the reference dataset to evaluate the loss of each received messenger to indicate the quality of its corresponding clients. Inter-client similarity, as another kind of auxiliary information, can also be deduced from the messengers. By calculating the Kullback--Leibler (KL) divergence between two messengers' label distributions, the central server quantifies the similarity between any two clients. Based on the client quality and inter-client similarity, the central server obtains and maintains $q$ dynamic high-quality candidates for each client, and each client only leverages the logits from its $k$ nearest neighbors among the candidates to improve its local model training. In other words, the central server dynamically creates and maintains a directed collaboration graph, in which a node only collaborates with its neighbors via messengers.  Compared with the existing heterogeneous FL paradigm,  SQMD does not require all clients to join the collaborative learning process simultaneously, as the low-quality clients (e.g., new clients) will not be selected as nearest neighbors to impact other clients. It should be noted here that any client, regardless of its quality, is assigned $k$ neighbors to improve its local training.  In addition, by limiting collaborations to a small number of quality neighbors,  SQMD significantly improves the personalization of client models and effectively alleviates global bias.

The major contributions are summarized as follows:
\begin{itemize}
\item We propose a new heterogeneous collaborative learning paradigm called SQMD, where clients can have heterogeneous model architectures to fully utilize the computing resources of each device and scale the collaboration range. These heterogeneous models collaborate with their nearest neighbors via logits (i.e., messengers) to protect their sensitive personal data and model parameters.
\item We design a novel collaboration protocol by exploiting two important kinds of auxiliary information in the messengers -- inter-client similarity and client model quality. This significantly enhances the personalization of client models and the robustness of the heterogeneous collaborative learning paradigm under asynchronous and sparse settings. To the best of our knowledge, this is the first work to exploit the logits' implicit information to build a dynamic collaboration graph.
\item We extensively evaluate our framework on three real-world datasets. The experimental results show that our SQMD achieves state-of-the-art classification accuracy. In addition, our SQMD is robust to the data sparsity and asynchronous settings that are pretty common in real-life healthcare applications.
\end{itemize}

\section{Related Work} \label{sec:background}
The communication graph empowers clients in decentralized learning \cite{kempe2003gossip} to establish a much more personalized connection. Many fully decentralized learning paradigms assume all devices communicate with others via short-range communication (e.g., Bluetooth). In other words, the communication graph is determined by the physical distance \cite{xie2019slsgd}. In this case, if a device has no physical neighbor (e.g., the distance to the nearest device exceeds the most extended communication range), it has to train the local model all alone. This can be a common phenomenon for some healthcare applications like rare chronic disease monitoring\cite{hurvitz2021establishing}. 
\hl{Some recent works} \cite{long2022decentralized, imran2022refrs, misra2016cross} \hl{propose to extract feature representations from model layers to generate dynamic communication graphs. However, models with alike behaviors are not necessarily parametrically approximate to each other. In this regard, we introduce semantic-level dynamic communication graphs in the FL paradigm to fully leverage the personalization potential in the messengers. The intuition is that models with alike behaviors on reference data will maintain their similarity while handling local data, which bypasses the impact from model parameters. Another core motivation of using logits instead of intermediate feature representations is that our learning process coordinates a variety of heterogeneous models, hence it is impractical to assume that two clients will have the same format (e.g., dimensionality and scale) for feature representations for similarity comparison. In contrast, using logits from the final layer facilitates comparisons across heterogeneous model structures via KL-divergence. In summary, logits are an alternative choice in such settings where comparing latent representations is infeasible.}

Knowledge distillation is initially introduced to take advantage of the experience from a pre-trained teacher model to improve the performance of student models with fewer model parameters \cite{phuong2019towards}. An unlabeled reference dataset will assist the knowledge transmission from the teacher to the student, which is also commonly adopted in semi-supervised learning \cite{xiaojin2008semi}. \cite{cheng2020explaining} and \cite{cho2019efficacy} provide the theoretical or empirical analytics of the knowledge distillation. Among these studies, the response-based knowledge distillation only needs to share the soft decisions (\hl{i.e.}, the outputs of the last fully-connected layer) for communications \cite{chen2017learning}. This is perfectly compatible with the heterogeneous device network where the classifiers have different model architectures. \cite{guo2020online} jointly trained some student models without any teacher model. \cite{zhang2021adversarial} further indicates that co-distillation between two student models may outperform the teacher-student hierarchy. 

Intuitively, for a personalized model of an individual in a homogeneous subset, the experience from the subset is more valuable than the corpora. Collaborative filtering (CF) \cite{su2009survey} is an excellent example. In typical CF methods, a group of users that are similar to the target user are regarded as its neighbors. The similarity is calculated based on users' explicit data (e.g., personal information and ratings on some items) or implicit data (e.g., purchase history, browsing history, and search patterns) \cite{schafer2007collaborative}. \cite{kleinberg2003convergent} shows that the experience from neighbors is theoretically helpful for personalized prediction. Such methods also work for the decentralized on-device learning framework \cite{he2018cola, koloskova2019decentralized, zhang2021democratic}. However, these frameworks either reveal the sensitive user data \cite{ye2022personalized} or uncover the model parameters \cite{he2018cola}. The proposed SQMD framework can solve these issues by introducing messenger distillation.

\begin{figure*}[t] 
\centering
\vspace{-12pt}
\includegraphics[scale=0.27]{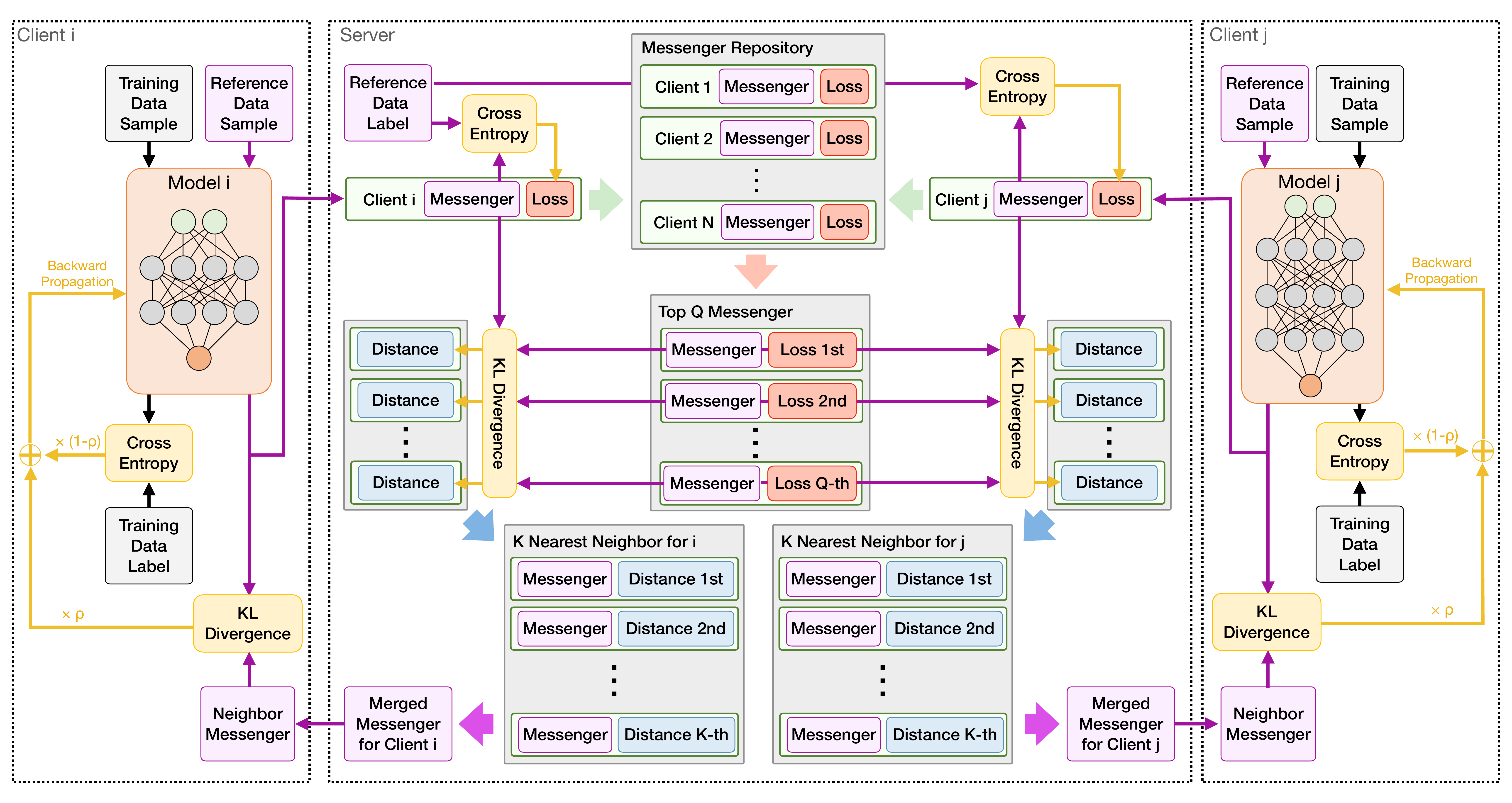}
\vspace{-20pt}
\caption{An overview of two clients with heterogeneous models and the server. The clients possess local samples, local training labels (grey for $\mathcal{D}_{\rm loc}^n$), and preloaded reference samples (purple for $\mathcal{D}_{\rm ref}$). $\mathcal{D}_{\rm ref}$ is identical across all clients. The client models will yield soft decisions for both training samples and reference samples. The latter one is the so-called messengers. During training, the client will send its messenger to the cloud server. Take the client $i$ for instance. The server first updates the central repository $\mathcal{S}$ with all received messengers and their loss and then selects $q$ high-quality (low-loss) messengers as $\mathcal{Q}$. Afterward, the server will calculate the KL divergence between messengers in $\mathcal{Q}$ and the messenger from client $i$. The nearest $k$ messengers in $\mathcal{Q}$ will be selected as the final $\mathcal{K}_i$. By adding the cross entropy of the training set and the KL divergence of the reference set (i.e., between the local messenger and the $\mathcal{K}_i$), client $i$ obtains a comprehensive loss for the backpropagation.}  
\label{Fig:model}
\vspace{-10pt}
\end{figure*}

\section{Methodology} \label{sec:method}
This section introduces the design and principle of the SQMD framework. We first provide preliminaries for some key components. Afterward, we formulate the research problem and demonstrate the proposed method.

\begin{table}[t]
\vspace{-12pt}
\caption{Notation table}
\begin{center}
\begin{tabular}{cc}
\hline
Notation & Definition \\
\hline 
$\mathcal{G} $ & The communication graph. \\
$\mathcal{A} $ & The client set. \\
$\mathcal{E} $ & The edges between clients. \\

$\mathcal{D}_{\rm loc}^n$ & The local dataset of the $n$-th user.   \\
$\mathcal{D}_{\rm ref}$ & The reference dataset. \\
$\mathcal{S}$ & The messenger repository in the server.   \\
$\mathcal{Q}$ & The set of high-quality messengers in $\mathcal{S}$.  \\
$\mathcal{K}^n$ & The neighbor set of the $n$-th client ($\mathcal{K}^n \in \mathcal{Q}$).  \\
$\Theta^n$ & The model parameter set of the $n$-th client.  \\

$\mathbf{C} $ & The dynamic weight matrix in $\mathcal{G}$.\\

$\bar{\mathbf{x}}_i/\textbf{x}_{i}^n$ & The $i$-th sample of $\mathcal{D}_{\rm ref}/\mathcal{D}_{\rm loc}^n$. \\
$\bar{\mathbf{y}}_i/ \mathbf{y}_i^n$ & The $i$-th label of $\mathcal{D}_{\rm ref}/\mathcal{D}_{\rm loc}^n$.\\
$\mathbf{s}^n$ & The messenger of the $n$-th client.\\
$\hat{\mathbf{s}}_i$  & The $i$-th messenger in $\mathcal{Q}$  \\
$\bar{\mathbf{s}}^n_i$ & The $i$-th messenger in $\mathcal{K}^n$ \\

$N$ & The number of clients ($|\mathcal{A}| =N$). \\
$R$ & The size of the reference dataset ($|\mathcal{D}_{\rm ref}|=R$). \\
$M_n$ & The size of the local dataset in the $n$-th client. \\
$g_n$ & The messenger quality of the $n$-th client. \\
$d_{nm}/c_{nm}$ & \makecell[c]{The distance/similarity between the $n$-th and\\ the $m$-th clients.}\\ 
$L_{\rm loc}^n/L_{\rm ref}^n$ & The local/ref loss of the $n$-th client model. \\
$q$ & The size of high-quality messenger set ($|\mathcal{Q}|=q$). \\
$k$ & The size of neighbor set ($|\mathcal{K}^1|=\cdots=|\mathcal{K}^n|=k$). \\
$\rho$ & A hyperparameter controls the impact from neighbors.\\
\hline
\end{tabular}
\label{tab:notation}
\end{center}
\vspace{-10pt}
\end{table}

\subsection{Preliminaries} \label{sec:Preliminaries}
\textbf{Definition 1: Reference Dataset.} Each client possesses an identical reference dataset $\mathcal{D}_{\rm ref} = \{\bar{\mathbf{x}}_i\}_{i=1}^R$ and the server possesses the ground-truth labels $\{\bar{\mathbf{y}}_i\}_{i=1}^R$. This reference dataset is independent of the local training dataset $\mathcal{D}_{\rm loc}^n$ of each client $n$. The reference dataset is easily accessible in most healthcare applications, e.g., using desensitized public benchmarks. 

\textbf{Definition 2: Messenger.} The client $n$ will produce soft decision of the reference dataset $\mathcal{D}_{\rm ref}$, i.e., the probability distribution over all classes $\mathbf{s}^n = \{\phi(\Theta^n, \bar{\mathbf{x}}_i)\}_{i=1}^R$ at each iteration. We name $\mathbf{s}^n$ as the messenger of client $n$. Devices can then exchange knowledge with each other via those messengers without revealing local training data or explicit model parameters. The server maintains a messenger repository $\mathcal{S}=\{ \mathbf{s}^n\}_{n=1}^N$ by collecting messengers from all clients.

\textbf{Definition 3: Model Quality.} Since the server possesses the ground-truth labels of the reference dataset, it can evaluate the quality of the received messengers and then accordingly grade the client:
\begin{eqnarray}
g_n = \sum_{i=1}^{R} \ell( \phi(\Theta^n, \bar{\mathbf{x}}_i) , \bar{\mathbf{y}}_i),
\end{eqnarray}
where $\ell(\cdot)$ is a loss function. Considering clients with relatively higher losses may imply malicious participants or new framework members, they are ruled out from the downstream communication steps. The remaining messengers are denoted as $\mathcal{Q} = \{\hat{\mathbf{s}}_1, \hat{\mathbf{s}}_2,..., \hat{\mathbf{s}}_q\}$, where $ \hat{\mathbf{s}}_i \in \mathcal{Q}$ indicates the messenger with the $i$-th ($i\leq q$) lowest loss.

\textbf{Definition 4: Inter-model Similarity.} Intuitively, a pair of clients with similar local training data tends to obtain closer on-device models. Consequently, they will produce convergent messengers. Hence we use the divergence between messengers to represent the affinity between clients:
\begin{eqnarray}
d_{nm} = \frac{1}{R}\sum_{j=1}^{R} D_{KL}( \mathbf{s}^n_j \| \mathbf{s}^m_j),
\end{eqnarray}
where $D_{KL}(\cdot)$ is the Kullback--Leibler divergence. We use $c_{nm}=\frac{1}{d_{nm}}$ to represent the similarity from devices $n$ to $m$. Notice that $c_{nm} \neq c_{mn}$. Both the model quality and inter-model similarity are used to rectify the communication neighbors of every device.

\textbf{Definition 5: Device Network.} 
The cloud server in SQMD will maintain a communication graph to coordinate client connections. Let $\mathcal{G} = (\mathcal{A}, \mathcal{E}, \mathbf{C})$ be a communication graph, where $\mathcal{A}$ is the client set with $N$ clients, and $\mathcal{E} \in \mathcal{A} \times \mathcal{A}$ indicates the set of edges between clients. $\mathbf{C}\in \mathbb{R}^{N \times N}$ is a dynamic weight matrix, where the weight of edge $(n, m)\in \mathcal{E}$ is denoted by $c_{nm} \in \mathbf{C}$ (i.e., inter-model similarity). The weight matrix $\mathbf{C}$ will be updated as the similarity between clients changes during training. 
We use $\mathcal{K}^n = \{\bar{\mathbf{s}}^n_1, \bar{\mathbf{s}}^n_2,...,\bar{\mathbf{s}}^n_k\}$, a subset of $\mathcal{Q}$, to indicate client $n$'s neighbor set, where $\bar{\mathbf{s}}^n_i\in \mathcal{K}^n$ is the messenger of the $i$-th ($i\leq k$) nearest neighbor (i.e., scores $i$-th highest $c_{ni}$) of $n$. Each client in the network will receive its $\mathcal{K}^n$ after $\mathcal{G}$ is updated.

\subsection{Problem Formulation}
Consider a personalized predictive task that involves $N$ user devices (clients). Each user device $n$ ($1\leq n \leq N$) possesses a private machine learning model $\phi(\cdot)$ parameterized by $\Theta^n$. Meanwhile, the $n$-th user device possesses a local dataset $\mathcal{D}_{\rm loc}^n = \{(\textbf{x}_{i}^n,\textbf{y}_{i}^n)\}_{i=1}^{M_n}$ that contains $M_n$ personal and sensitive data samples. $\textbf{x}_{i}^n$ denotes the input feature vector and $\textbf{y}_{i}^n$ is a one-hot vector over all classes representing the corresponding ground truth, e.g., medical records paired with diagnosed diseases. Our goal is to learn a function that maps every input data point to the correct class. We define the local objective function $L_{\rm loc}$ of a single user device $n$ w.r.t. $\mathcal{D}_{\rm loc}^n$ as:
\begin{eqnarray}\label{eq:local_loss}
L_{\rm loc}^n = \sum_{i=1}^{M_n} \ell (\phi(\Theta^n,\mathbf{x}_i^n), \mathbf{y}_i^n),
\end{eqnarray}
where $\ell$ is a loss function (i.e., cross-entropy in our case) that quantifies the classification error.

In a straightforward setting where each on-device model is optimized in isolation, the only training resource for device $n$ is the local dataset. So for each device we wish to minimize the gap between $\phi(\Theta_i^n, \mathbf{x}_{i}^n)$ and $\mathbf{y}_{i}^n$. Apparently, the learning objective of the whole network will be:
\begin{eqnarray}\label{eq:fully_decentralization}
\min_{\Theta^1, ... , \Theta^N}\sum_{n =1}^{N} L_{\rm loc}^n.
\end{eqnarray}
When the volume of every local dataset $M_n$ is big enough for training an independent and satisfactory client model, Eq.\ref{eq:fully_decentralization} can be achieved for both individuals and the population.

\begin{algorithm*}[htbp]
\caption{Similarity-\hl{Quality}-Based Messenger Distillation}
\label{Ag:1}
\textbf{Initialization}: Let $t=0$ be the index of iterations, $I$ be the communication interval. Let $\Theta_0^n$ be the initial variables of client $n$, $s_0^n=\{\phi(\Theta^n,\bar{\mathbf{x}}_i)\}_{i=1}^R$ be the initial messenger \hl{w.r.t.} reference dataset from client $n$. Let $\eta_t$ be the learning rate. 
\begin{algorithmic}[1]
\Repeat 
\State $t=t+1$
\For{$n=1,\cdots,N$}
\If {$t$ mod $I=0$}
\State The client $n$ generates messenger $\mathbf{s}_{t-1}^n \leftarrow \{\phi(\Theta_{t-1}^n,\bar{\mathbf{x}}_i)\}_{i=1}^R$
\State The client $n$ sends messenger $\mathbf{s}_{t-1}^n$ to the server
\State The server calculates the client model quality $g^n_t$ based on $\mathbf{s}_{t-1}^n$ and update messenger repository $\mathcal{S}_t$ 
\State The server selects $q$ messengers as $\mathcal{Q}_t$ based on model quality.
\State The server calculates inter-model similarity and selects $k$ nearest messengers from $\mathcal{Q}_t$ as $\mathcal{K}^n_t$.
\State The server sends $\mathcal{K}^n_t$ to client $n$
\EndIf
\State $\Theta_{t}^n \leftarrow \Theta_{t-1}^n - \frac{(1-\rho)\eta_t}{M_n}\sum_{(\mathbf{x}, \mathbf{y})\in D_n} \nabla \ell(\phi(\Theta_{t-1}^n,\mathbf{x}), \mathbf{y})- \frac{2\eta_t\rho}{R}\sum_{\bar{\mathbf{x}}\in D_r}(\nabla \phi (\Theta_{t-1}^n,\bar{\mathbf{x}}))^T(\phi(\Theta_{t-1}^n,\bar{\mathbf{x}})-\frac{1}{k}\sum_{m =1}^k \mathbf{s}_{t-1}^m))$
\EndFor
\Until{termination condition satisfied}
\end{algorithmic}
\end{algorithm*}

\subsection{Messengers-based Heterogeneous Collaboration} \label{sec:Objective}
In personalized prediction scenarios, the amount of data on each user device is limited, e.g., most e-health users do not have a long and diverse list of medical records for learning an automated diagnosis model. Meanwhile, a fully isolated model is prone to unreliable signals and noises if deployed on IoT sensors. 
 These two main obstacles hold the isolated paradigm back from the perspective of practicality. On the contrary, centralized learning paradigms can leverage data gathered from all users to learn a global model, thus minimizing the negative impact of data insufficiency and outliers. But such non-personalized models usually sacrifice the correctness of the minority to obtain better overall accuracy. Furthermore, this amplifies privacy concerns and incurs a high demand for resource-intensive computing facilities. 

Though conventional gradient-sharing or weight-sharing methods seem to be a workable solution, they unfortunately hold the assumption of model homogeneity, which is \hl{increasingly} impractical given the diversity of modern personal devices and their varying computational capacities. In light of this, we propose a similarity-quality-based semi-decentralized information exchanging protocol, enabling collaboration among devices when learning heterogeneous personalized models. Unlike existing solutions, the clients in SQMD merely need to share their messengers on a reference dataset, drastically reducing the risk of privacy breaches for both data and models. The messengers can be also viewed as a compact yet informative summary of each client model that requires far less communication bandwidth than transmitting models/gradients. Besides, to establish the personalization of the client model, each client only communicates with $k$ high-quality neighbors, i.e., a small alterable group of devices $\mathcal{K}^n$ selected from set $\mathcal{Q}$. For client $n$, by utilizing the reference dataset $\mathcal{D}_{\rm ref}$, we quantify the disagreement between it and its $k$ neighbors in $\mathcal{K}^n$:
\begin{align}\label{eq:ref_loss}
L_{\rm ref}^n=\sum_{j=1}^R \Big\| \phi(\Theta^n,\bar{\mathbf{x}}_j)-\frac{1}{k}\sum_{m\in \mathcal{K}^n} \phi({\Theta}^m,\bar{\mathbf{x}}_j)\Big\|^2,
\end{align}
where $\phi(\Theta^n,\bar{\mathbf{x}}_j)=\mathbf{s}^n$ is the soft decisions from the local device $n$ {w.r.t.} the $j$-th reference sample $\bar{\mathbf{x}}_j \in \mathcal{D}_{\rm ref}$, and $\frac{1}{k} \sum_{m\in \mathcal{K}^n} \phi(\Theta^m,\bar{\mathbf{x}}_j)$ is the ensemble of $k$ neighbors' messengers on the same sample.

We hereby propose the overall objective for training all distributed devices in the network in SQMD:
\begin{align}\label{eq:d_train_obj}
L^*=\min_{\Theta^1, ... , \Theta^N}\sum_{n =1}^N \bigg( (1-\rho) L_{\rm loc}^n + \rho~L_{\rm ref}^n\bigg),
\end{align}
where $\rho$ is a trade-off hyperparameter. Essentially, solving this equation means that the algorithm converges to the following set of knowledge distillation stationary points:
\begin{align}\label{eq:d_stationary_pts}
\mathcal{A}^* =& \Bigg\{\Theta \bigg| \nabla_{\Theta^n}\bigg(\sum_{i=1}^{M_n} \ell (\phi(\Theta^n,\mathbf{x}_i^n), \mathbf{y}_i^n)+ \rho\sum_{j=1}^R \Big\| \phi(\Theta^n,\bar{\mathbf{x}}_j) - \nonumber \\ 
&\frac{1}{k}\sum_{m\in \mathcal{K}^n} \phi({\Theta}^m,\bar{\mathbf{x}}_j)\Big\|^2\bigg)=0, \forall n \Bigg\}.
\end{align}
In this way, SQMD requires all devices to reach for the optimum on their local data and the consensus on the reference data with their neighbors simultaneously. With Eq.~\ref{eq:ref_loss}, SQMD allows knowledge from peer models to be passed through via the soft decisions on $\mathcal{D}_{\rm ref}$, without any assumption of homogeneous model architectures.

\subsection{Overview and Algorithm of SQMD}\label{sec:Algorithm}
The SQMD is presented in detail in Algorithm \ref{Ag:1}. 
To reach the stationary status described in Eq.~\ref{eq:d_stationary_pts}, SQMD iteratively minimizes the local loss as well as the disagreement with neighbors for each client. Lines $5-10$ are the communication step. 
At each iteration $t$, client $n$ firstly generates its own messenger w.r.t. the shared reference dataset with its newest model and sends it to the cloud server. 
After that, the server will update the central repository $\mathcal{S}_t$ and the high-quality messenger set $\mathcal{Q}_t$. The $k$ most similar messengers in $\mathcal{Q}_t$ will be sent to the client $n$. 
We provide a device-level view on SQMD in Fig.~\ref{Fig:model}, where the core similarity- and quality-based filtration steps are conducted in the server. Then, line $12$ is the model update step where client $n$ computes new parameters with the help of knowledge from its neighbors. 

\subsection{Framework Applicability}\label{sec:Universality}
\hl{The default setting of the communication interval $I$ is $1$ in} Algorithm \ref{Ag:1}. \hl{In other words, all clients send their latest messenger to the server as soon as their local models are updated. For some small-scale IoT collaboration applications, it would be more cost-effective to apply a longer communication interval. This may slow down the convergence rate but will not significantly influence the eventual performance.
In terms of the local model, the proposed SQMD framework does not have any assumptions on the model structure of each client. It supports collaboration between any models with heterogeneous designs, including Multilayer Perceptron (MLP), Convolutional Neural Network (CNN), and Recurrent Neural Network (RNN), as long as the output dimensionality is the same.
Notably, the reference dataset is supposed to be balanced in class distribution. Data augmentation techniques} \cite{yu2021socially} \hl{are recommended if the original reference samples of different classes are not of the same order of magnitude. As such, the quality ranking and neighbor selection results generated by the server will be guaranteed.}

\section{Experiment}  \label{sec:experiment}
In this section, extensive experiments are conducted on real-world datasets to evaluate the feasibility of our SQMD in three on-device healthcare classification tasks. Specifically, the following four research questions (\textit{RQs}) are studied:

\begin{enumerate}
\item [\textit{RQ1:}] How is the performance competitiveness of our method compared with the baselines?
\item [\textit{RQ2:}] How robust can SQMD be while increasing the sparsity of local training data?
\item [\textit{RQ3:}] How do the hyperparameters influence our SQMD when performing real-life e-health analytics?
\item [\textit{RQ4:}] How stable is SQMD in the asychonorous on-device training?

\end{enumerate}

\vspace{-10pt}
\subsection{Datasets and Baselines} \label{sec:datasets}
Three real-life datasets are used in the experiments, where the first one is \textbf{Sleep Cassette (SC)} database \cite{mourtazaev1995age}, which includes $153$ overnight polysomnography (PSG) recordings. We extract $40$ clear Electroencephalogram (EEG) records from the PSG for sleep quality rating. Three labels, \hl{i.e.}, awake, non-rapid eye movement sleep (NREM), and rapid eye movement sleep (REM) are derived from the original Rechtschaffen and Kales sleep stage annotations \cite{rechtschaffen1968manual} for the classification task. 

The second dataset is \textbf{PhysioNet Apnea-ECG Dataset (PAD)} \cite{ichimaru1999development}. It is a famous dataset regarding a common sleep-related disease -- Obstructive Sleep Apnea (OSA). PAD contains $35$ overnight ECG recordings from severe patients, moderate patients, and normal people. The ECG signals in PAD are converted into RR interval (pulse-to-pulse interval) signals via the algorithm provided by \cite{cai2020qrs}. Specifically, the model will tell whether there is an apnea event based on a $60$-dimensional RR-interval (time interval between two R-peak) vectors. Such classification pipelines greatly facilitate OSA patients' self-monitoring since they can instantly get detection results via wireless wearable devices like smartwatches and smart-bands\cite{ye2021fenet}. 

Besides, we further conduct experiments on  \textbf{\hl{Fashion-MNIST} (FMNIST)} \cite{xiao2017fashion} to support additional comparisons. It has been widely adopted to evaluate the model performance in benchmarking distributed machine learning algorithms. Since there is no explicit relationship between samples for grouping (e.g., samples from one individual can be regarded as one device in SC and PAD), we conduct a random and even segmentation on FMNIST, which is proposed by \cite{bistritz2020distributed}, to make it compatible with distributed on-device learning scenarios.

We compare our SQMD with the following three distributed optimization strategies:
\begin{itemize}
\item \textbf{FedMD}\cite{li2019fedmd}: A heterogenous federated learning framework using model distillation. The communication is conducted based on an unlabeled reference dataset. All participants will send their logits to a central server and receive an average one as the global knowledge. FedMD can be regarded as a simplified case of SQMD that $q=k=|\mathcal{A}|$. 
\item \textbf{D-Dist}\cite{bistritz2020distributed}: A decentralized distillation framework. The inter-device communication also relies on an unlabeled reference dataset, but the central server is removed. Each device in the network will communicate with a static group of peers at each iteration. The groups are randomly sampled from the network. Notice that when the size of all groups is as large as the whole network $|\mathcal{A}|$, D-Dist is equivalent to FedMD.
\item \textbf{I-SGD}\cite{lian2017can}: Every device in the network has an independent model and optimizes the parameters without cross-device collaborations.
\end{itemize}

\begin{table}[t]
\vspace{-12pt}
\caption{Statistics of experimental datasets.}
\begin{center}
\begin{tabular}{cccc}
\hline
Dataset & SC & PAD & FMNIST \\
\hline
\#Clients & $32$ & $28$ & $20$ \\  
\#ResNet8& $10$ & $9$ & $6$ \\ 
\#ResNet20& $11$ & $9$ & $7$ \\ 
\#ResNet50& $11$ & $10$ & $7$ \\ 
\#Sample & $158,301$ & $131,917$ & $70,000$ \\
\#Class & $3$ & $2$ & $10$ \\  
\hline
\end{tabular}
\label{tab:datasets}
\end{center}
\vspace{-20pt}
\end{table}

\vspace{-8pt}
\subsection{Experimental Setting}\label{sec:Setting}
In SC and PAD, each recording of a patient is regarded as one slice, so they have $40$ and $35$ slices, respectively. Sliding window data augmentation is adopted to generate more samples on each slice in these two datasets. We randomly and evenly divide training samples in the FMNIST into $20$ slices. Considering the data is \hl{i.i.d.} across slices in FMNIST, we remove training samples belonging to one random class from each slice.
For SC and PAD, $20\%$ of the entire slices are randomly selected and combined as the reference dataset, and the rest of them are regarded as the local datasets of separate clients. For FMNIST, the original $10000$ testing samples are used as the reference dataset. That is to say,  $N_{\rm SC}=32$, $N_{\rm PAD}=28$, $N_{\rm FMNIST}=20$. We then randomly split the samples in each local dataset for training, validation, and testing with a ratio of 8:1:1. 

Both SQMD and all baselines are versatile frameworks that support the joint optimization of arbitrary heterogeneous models. To evaluate the performance of the SQMD framework and the baselines, we use the ResNet \cite{he2016deep} with different numbers of layers (i.e., ResNet8, ResNet20, ResNet50) as benchmark heterogeneous DNN models. For the SC and PAD datasets (time series), all the 2D-convolutional layers in the ResNet are replaced with 1D-convolutional layers. \hl{The reason we chose ResNet is that it is a well-established and versatile model across various tasks, with a desirable balance between performance and efficiency. Additionally, ResNet is also used by D-Dist and I-SGD baselines} \cite{bistritz2020distributed}\hl{, which allows for a fair horizontal comparison.} The local loss $\ell$ is specified as cross-entropy for all devices as it is commonly used in such classification tasks. \hl{The communication interval is set to the default value, i.e., $I=1$}. The primary statistics of the preprocessed datasets and the model ratio configuration are shown in Table~\ref{tab:datasets}.

Accuracy (Acc), precision (Pre), \hl{and} recall (Rec) are adopted to evaluate the performance of SQMD and three baselines on all three datasets (macro-precision and macro-recall for multi-class datasets). The hyperparameter tuning was conducted via grid search. The optimal values and search intervals are listed in Table ~\ref{tab:hyperparameters}.

\begin{table}[t]
\vspace{-12pt}
\caption{Hyperparameter settings.}
\begin{center}
\begin{tabular}{cccc}
\hline
Dataset & Hyperparameter & \makecell[c]{Optimal \\ Value } & Search Interval \\

\hline
\multirow{2}{*}{SC}
& $q$ & $16$ & $\{4,8,12,16\}$  \\
& $k$ & $0.5q$ & $\{0.25q, 0.5q, 0.75q, q\}$  \\  
& $\rho$ & $0.8$ & $\{0.1,0.4,0.6,0.8\}$ \\
\hline
\multirow{2}{*}{PAD}
& $q$ & $12$ & $\{4,8,12,16\}$  \\
& $k$ & $0.5q$ & $\{0.25q, 0.5q, 0.75q, q\}$ \\  
& $\rho$ & $0.8$ & $\{0.1,0.4,0.6,0.8\}$ \\
\hline
\multirow{2}{*}{FMNIST}
& $q$ & $16$ & $\{4,8,12,16\}$  \\
& $k$ & $0.75q$ & $\{0.25q, 0.5Qq, 0.75q, q\}$  \\  
& $\rho$ & $0.8$ & $\{0.1,0.4,0.6,0.8\}$ \\
\hline
\end{tabular}
\label{tab:hyperparameters}
\end{center}
\vspace{-15pt}
\end{table}

\subsection{Effectiveness Discussion  (\textit{RQ1})} \label{sec:Effectiveness}
After evaluating the classification performance of our SQMD and three baselines, the average test accuracy and standard deviations of five independent experiments are listed in Table \ref{tab:effectiveness}. We draw the following observations from the experimental results. 

Firstly, SQMD consistently outperforms all baselines on all datasets in terms of all metrics. This validates the efficacy of similarity- and quality-based collaborative learning in the proposed SQMD framework. Secondly, though the I-SGD achieves the lowest accuracy on the FMNIST, it outperforms FedMD and D-Dist methods on two healthcare datasets. In general, introducing inter-device communication will enhance the learning ability of distributed models. The results on the FMNIST dataset are in line with this intuition. The unusual results on SC and PAD suggest that the knowledge from neighbors without careful selection can be deleterious to local models. Additional experiments are conducted in Section \ref{sec:Sparsity} to further investigate this observation. Lastly, the results imply that the SQMD surpasses FedMD, which is also an unusual phenomenon. Since the devices in FedMD will receive knowledge from all other devices ($k=N-1$) during training, FedMD is the theoretic skyline for both SQMD and D-Dist ($k<N-1$). A possible explanation is that some devices are more `beneficial' to the local model and some are more `detrimental'. If peers in the framework are randomly selected as neighbors (D-Dist), the skyline would be FedMD. But if only `beneficial' ones are selected (SQMD), then the performance may exceed the FedMD. This assumption is verified in Section \ref{sec:Hyperparameters}.

\begin{table}[t]
\vspace{-0pt}
\caption{Performance of SQMD and baselines on three datasets}
\begin{center}
\begin{tabular}{c|c|c|c|c|c}
\hline
Method & Metric & SQMD & FedMD & D-Dist & I-SGD  \\
\hline 
\multirow{3}{*}{SC}
& Acc  & \textbf{0.8465} & 0.8434 & 0.8366 & 0.8449 \\
& Pre & \textbf{0.7547} & 0.7518 & 0.7444 & 0.7532\\
& Rec &  \textbf{0.8056} & 0.8043 & 0.7946 & 0.8048\\
\hline
\multirow{3}{*}{PAD} 
& Acc  & \textbf{0.9203} & 0.9152 & 0.9072 & 0.9163\\
& Pre  & \textbf{0.8966} & 0.8889 & 0.8850 & 0.8922\\
& Rec  & \textbf{0.8861} & 0.8717 & 0.8601 & 0.8793\\
\hline
\multirow{3}{*}{FMNIST}
& Acc  & \textbf{0.8738} & 0.8678 & 0.8650 & 0.8629 \\
& Pre & \textbf{0.8761} & 0.8693 & 0.8666 & 0.8680\\
& Rec & \textbf{0.8727} & 0.8667 & 0.8640 & 0.8644\\
\hline
\end{tabular}
\label{tab:effectiveness}
\end{center}
\vspace{-20pt}
\end{table}

\subsection{Robustness to Data Sparsity (\textit{RQ2})}  \label{sec:Sparsity} 
\begin{figure}[t]
\centering
\vspace{-22pt}
\subfigure[]{
\begin{minipage}[t]{0.48\linewidth}
\centering
\includegraphics[scale=0.49]{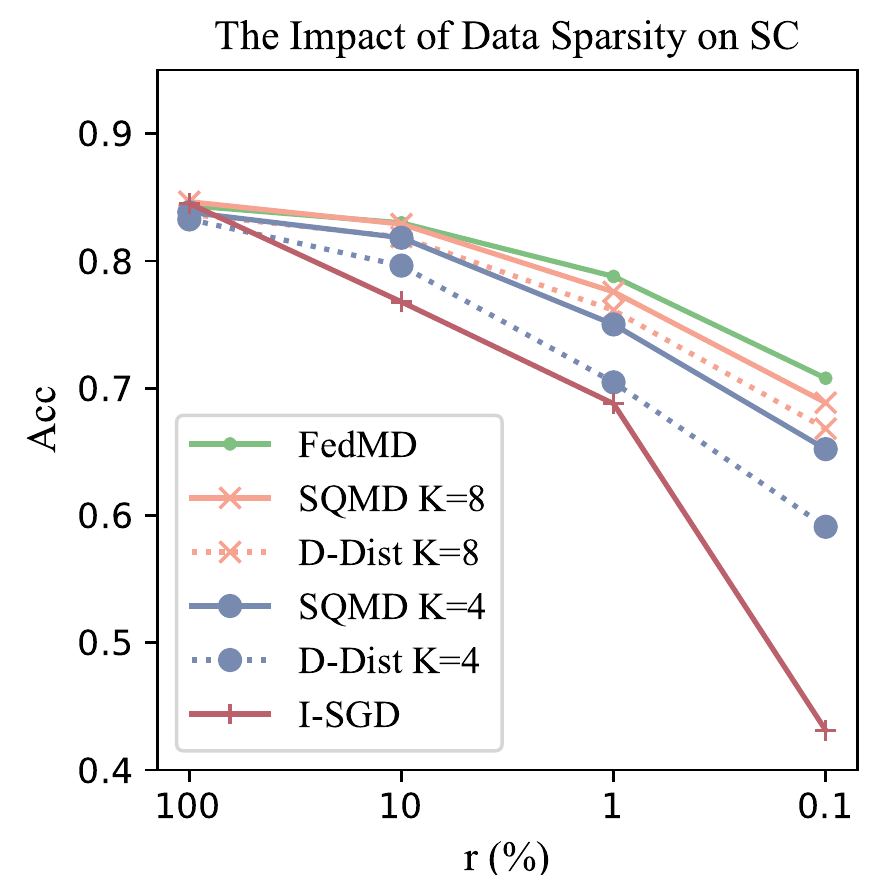}
\end{minipage}%
\label{fig:sparsity1}
}%
\subfigure[]{
\begin{minipage}[t]{0.48\linewidth}
\centering
\includegraphics[scale=0.49]{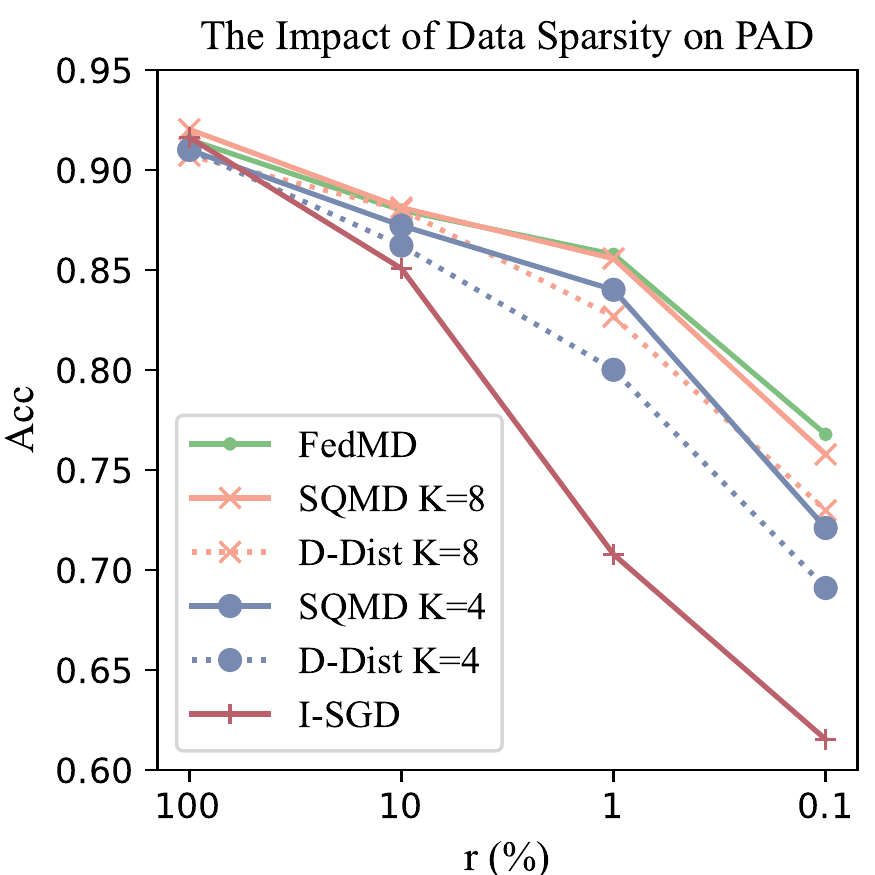}
\end{minipage}%
\label{fig:sparsity2}
}%
\centering
\vspace{-7pt}
\caption{(a) and (b) depict the performance of SQMD with different numbers of neighbors (i.e., SQMD $k=4$ and SQMD $k=8$) and baselines on SC and PAD when the data sparsity $r$ goes down. The I-SGD is the worst method on both datasets when $r \leq 10\%$. Our SQMD constantly outperforms D-Dist while having the same number of neighbors.}
\label{fig:e_sparsity}
\vspace{-12pt}
\end{figure}

We answer RQ2 by investigating the performance of the proposed framework in a set of sparse data training simulations. The simulated data sparsity is implemented by randomly extracting $r$\% of samples in SC and PAD datasets. Tested network communication strategies include FedMD, I-SGD, SQMD (with $4$ and $8$ neighbors), and D-Dist (with $4$ and $8$ neighbors). The test accuracy is depicted in Figure \ref{fig:e_sparsity}. We have similar observations in other metrics and do not report them due to space limitations. When $r$ decreases from $100$\% to $0.1$\%, all the methods suffer \hl{from} a distinct performance decline, especially I-SGD. Apparently, as less data is available for the training processes, the over-fitting issue becomes more serious, which significantly hurts the model's generalization. We also notice that increasing the number of neighbors $k$ will remarkably increase the model robustness to different levels of data sparsity. The largest gains on SC and PAD are $64.15$\% and $24.78$\%, respectively.
In general, collaborating with more neighbors can achieve higher accuracy, except in one special case. SQMD ($k=4$) outperforms D-Dist ($k=8$) on PAD when $r=1$. It implies that the divergence between selective collaboration (SQMD) and random collaboration (D-Dist) is more obvious with the decreasing availability of training data. In other words, the local model is more fragile confronting `detrimental' neighbors. To be specific, the mean improvements resulted from replacing random connection with similarity-quality-based connection when $r=10$, $1$, and $0.1$ are $1.08$\%, $3.23$\%, and $3.49$\%, respectively. 

In a nutshell, inter-device communication allows each device to get more information from each other when local data is limited, which empowers distributed distillation methods to resist data sparsity. Moreover, distributed distillation methods with similarity-quality-based neighbor filtration, our SQMD, can further strengthen this kind of robustness.

\begin{figure*}[htbp]
\centering
\vspace{-12pt}
\subfigure[]{
\begin{minipage}[t]{0.23\linewidth}
\centering
\includegraphics[scale=0.49]{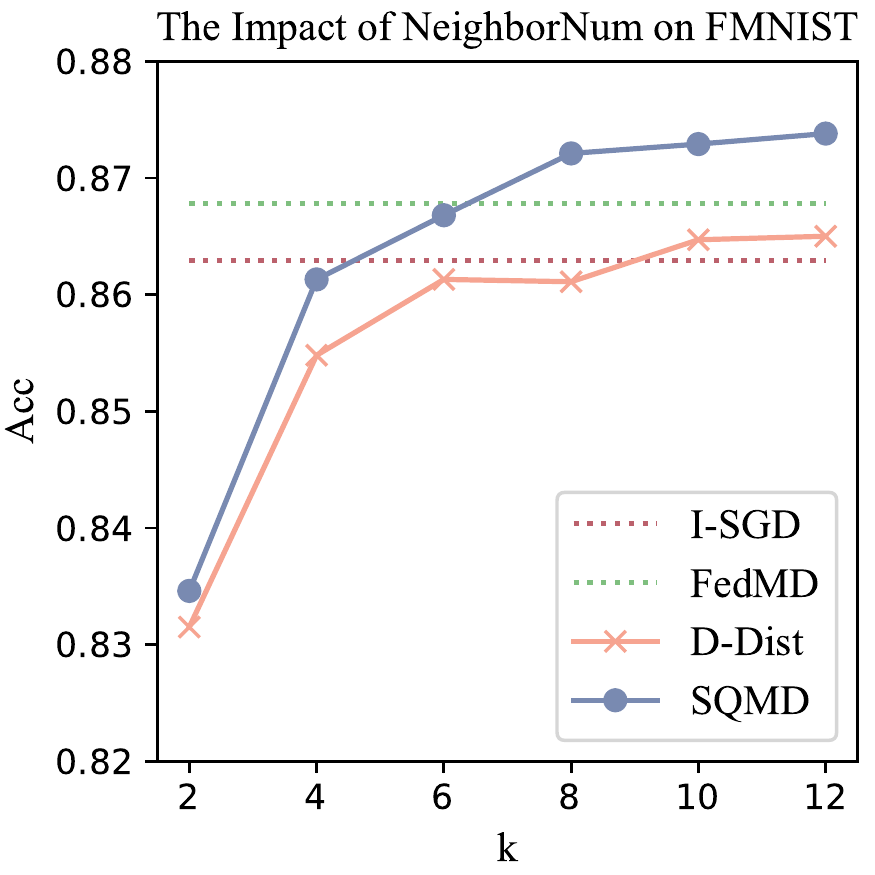}
\end{minipage}%
\label{fig:hyper_K_MN}
}%
\subfigure[]{
\begin{minipage}[t]{0.23\linewidth}
\centering
\includegraphics[scale=0.49]{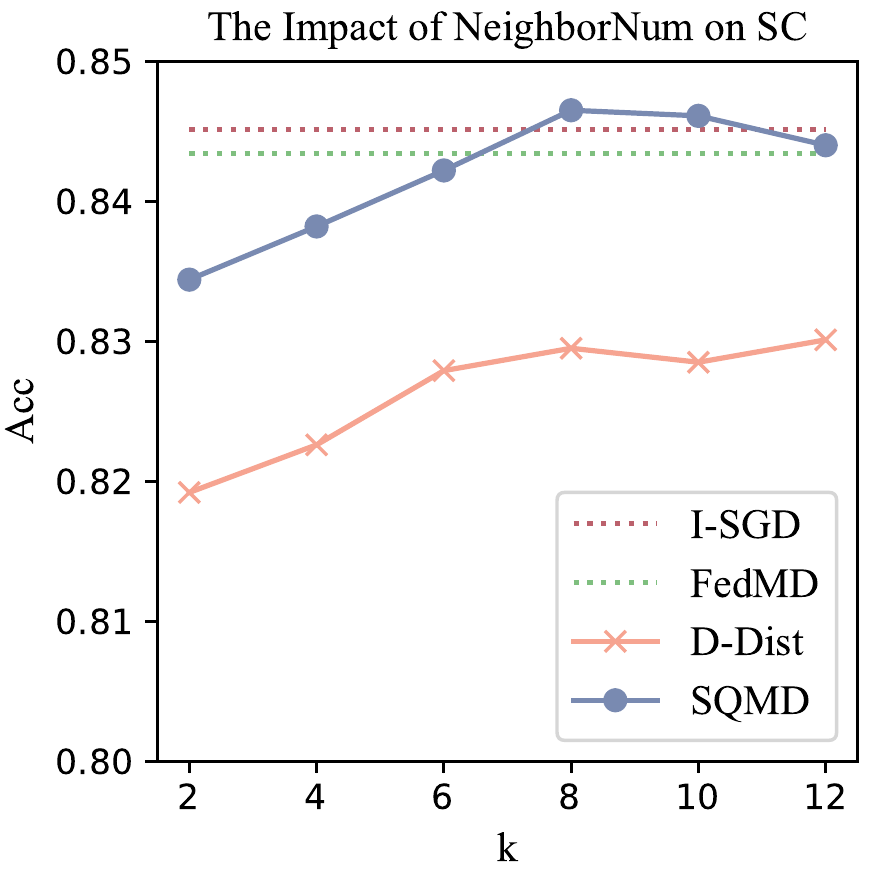}
\end{minipage}%
\label{fig:hyper_K_SC}
}%
\centering
\subfigure[]{
\begin{minipage}[t]{0.23\linewidth}
\centering
\includegraphics[scale=0.49]{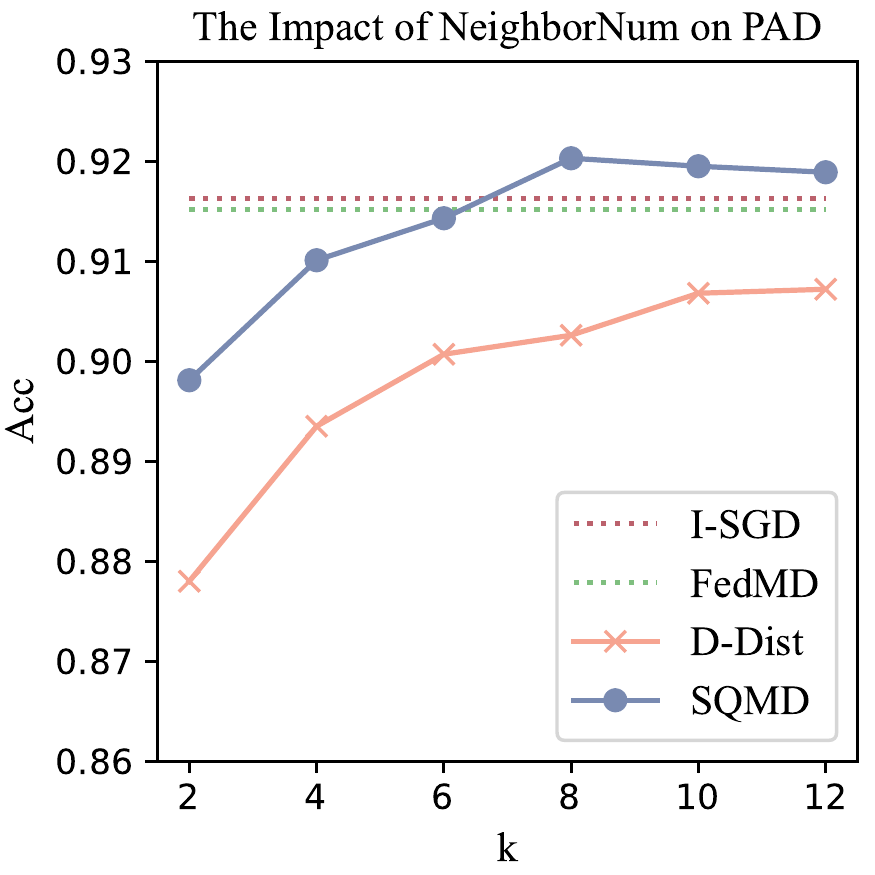}
\end{minipage}%
\label{fig:hyper_K_PAD}
}%
\subfigure[]{
\begin{minipage}[t]{0.23\linewidth}
\centering
\includegraphics[scale=0.49]{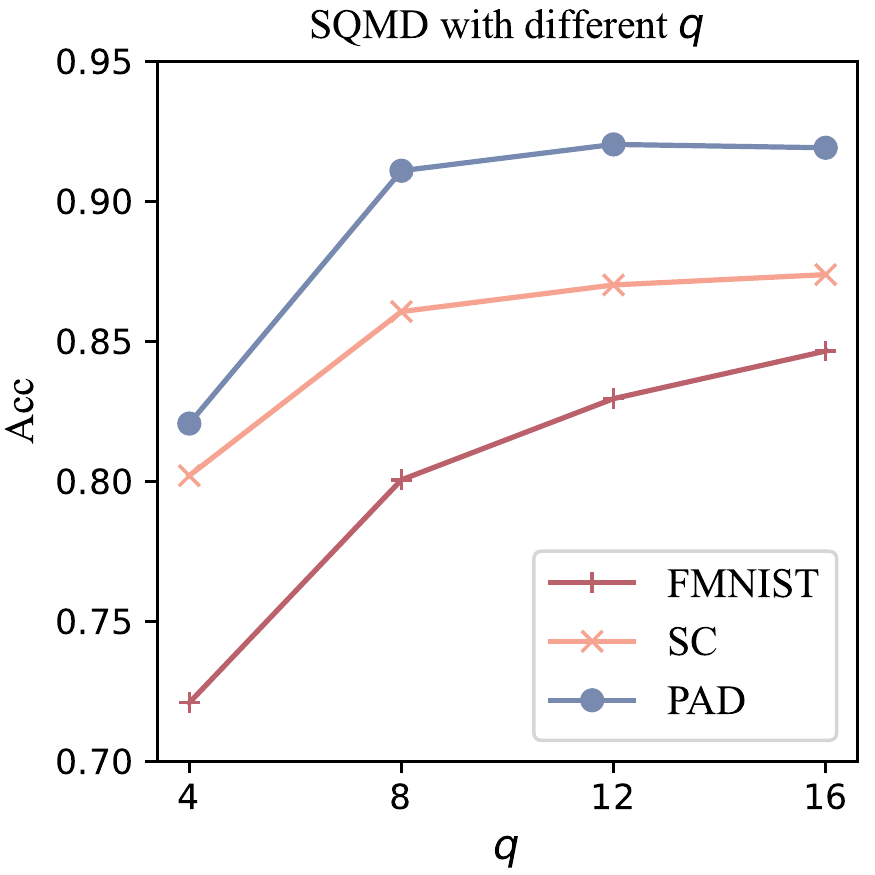}
\end{minipage}%
\label{fig:hyper_Q}
}
\vspace{-5pt}
\caption{The performance of SQMD and baselines on FMNIST (a), SC (b), and PAD (c) with different numbers of neighbors $k$, and the performance of SQMD on three datasets with different values of $q$ (d).}
\label{fig:hyper}
\vspace{-12pt}
\end{figure*} 

\subsection{Hyperparameter Sensitivity (\textit{RQ3})}  \label{sec:Hyperparameters}
In this section, we showcase the effect of two hyperparameters, namely $k$ and $q$, on the performance of SQMD.

Firstly, we enlarge $k$ from $2$ to $12$ while $q$ follows the optimal setting in this Section~\ref{sec:Effectiveness}. We illustrate the outcomes in Fig.~\ref{fig:hyper_K_MN}-\ref{fig:hyper_K_PAD}. The accuracy figures of I-SGD and FedMD are also presented as a reference line for $k=0$ and $k=N-1$. The classification performance of D-Dist and SQMD on FMNIST constantly goes up as $k$ increases. This validates that distributed distillation can \hl{transfer} knowledge across devices. D-Dist tends to converge to the green dotted line (FedMD), which is in line with our intuition since FedMD is equal to D-Dist when each participant in D-Dist has $N-1$ neighbors. 
The counter-intuitive observation is that SQMD outperforms the FedMD on all three datasets. We try to explain this phenomenon with an example. Imagine an extreme scenario where four devices, namely A to D, compose a network. A and B only have negative samples, while C and D only have positive samples for binary classification. The models on all devices can reach $100$\% accuracy by training locally only. It is conceivable that introducing fully inter-device communication will worsen the overall accuracy ($< 100$\%). However, if every device in the network only communicates with the most similar one (i.e., A communicates with B, and C communicates with D), the overall accuracy remains unchanged. This reveals how different communication/collaboration strategies influence the performance of distributed on-device learning algorithms with different data distributions on different client devices.
Note that I-SGD is superior to FedMD on both SC and PAD datasets because the local data distributions of some SC and PAD clients significantly differ from their global distributions.  In this case, the global knowledge (FedMD) or a random segment of the global knowledge (D-Dist) will worsen the performance of these clients.
Next, we show the test accuracy of SQMD on all datasets in Fig.~\ref{fig:hyper_Q}  when $k=0.5q$ and $q$ vary in $\{4,8,12,16\}$. The results suggest that the best performance on PAD is obtained when $q$ is between $8$ and $16$. Meanwhile, higher $q$ leads to more obvious performance improvement on FMNIST. This suggests that a network with similar data distribution benefits more directly and swiftly from inter-device information sharing.

\begin{figure}[hbtp]
\flushleft
\vspace{-10pt}
\subfigure[]{
\begin{minipage}[t]{1\linewidth}
\flushleft
\includegraphics[scale=0.43]{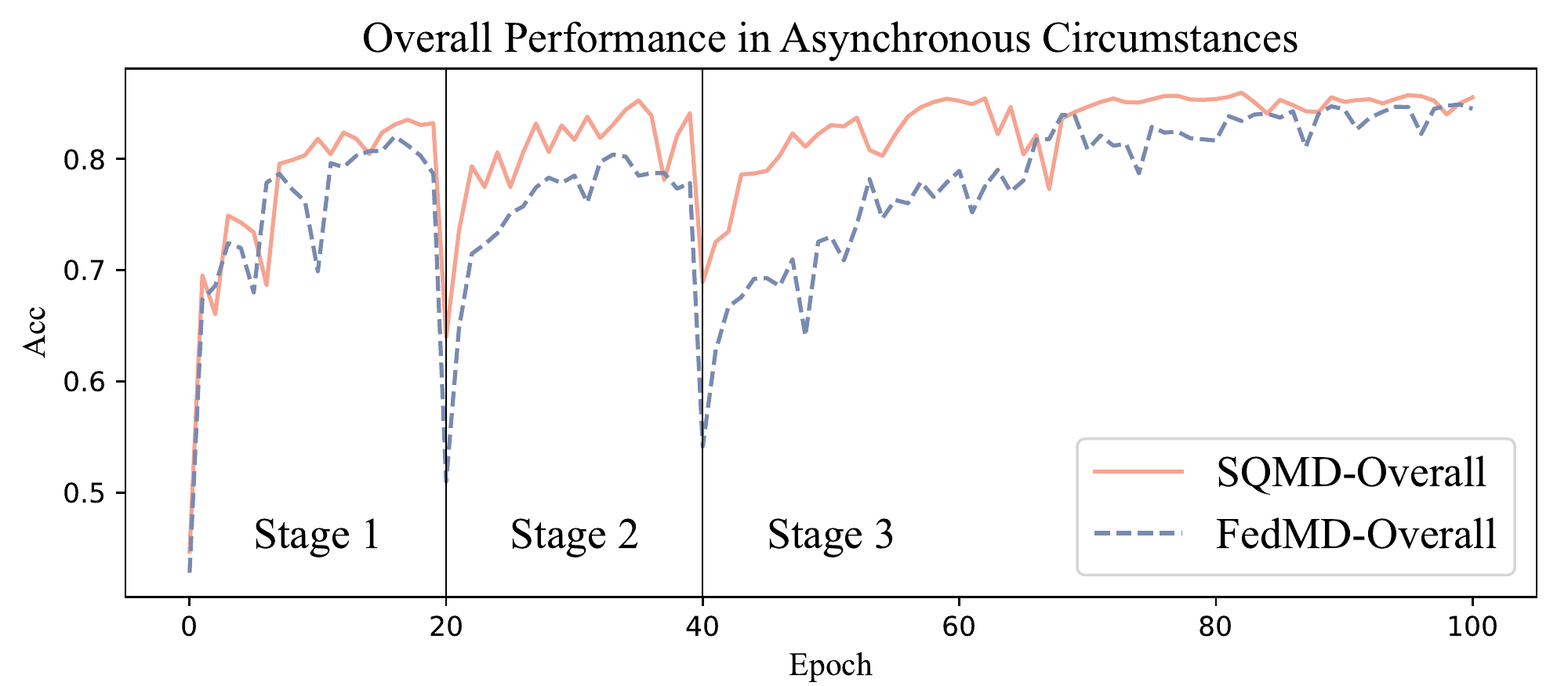}
\end{minipage}%
\label{Fig:case_study_1}
}%
\vspace{0cm}
\subfigure[]{
\begin{minipage}[t]{1\linewidth}
\flushleft
\includegraphics[scale=0.43]{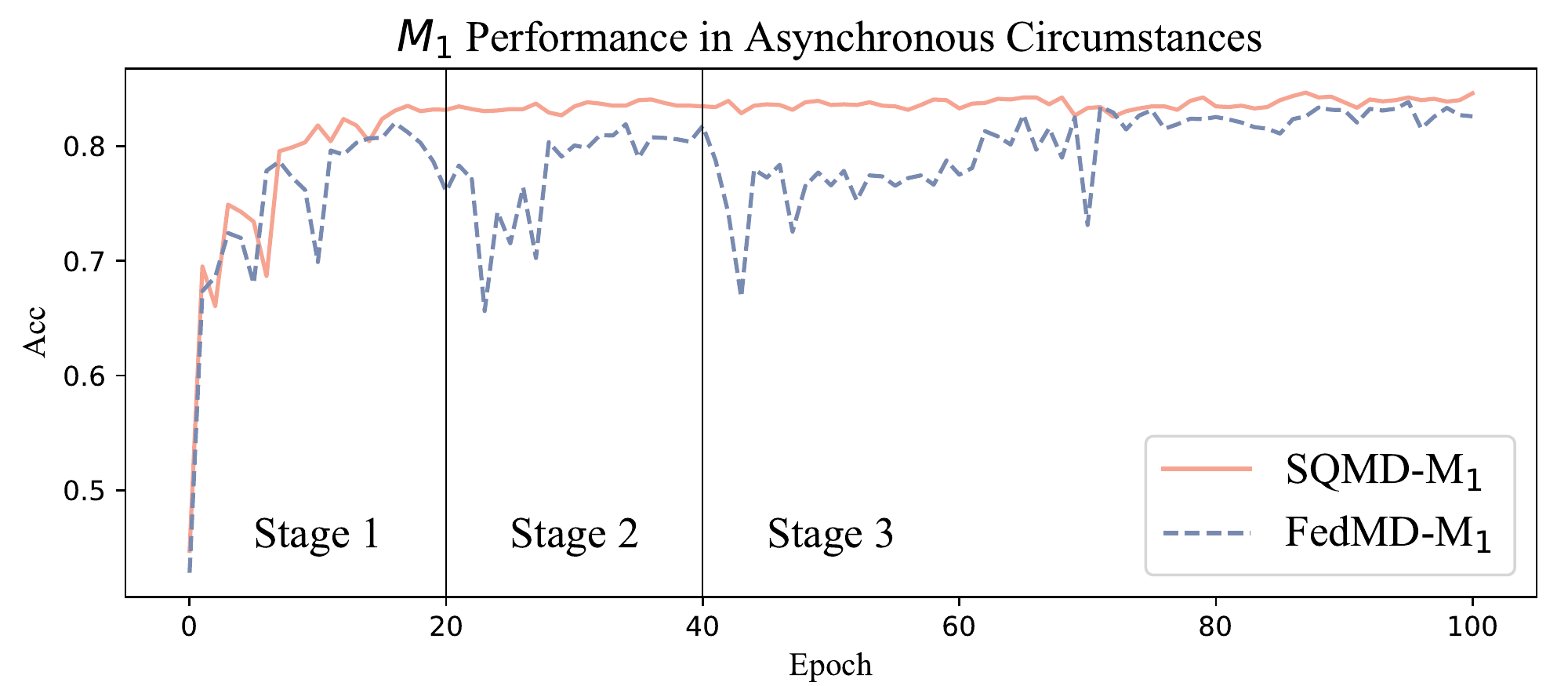}
\end{minipage}%
\label{Fig:case_study_2}
}%
\vspace{-0.18cm}
\caption{The overall test accuracy (a) and partial ($M_1$) test accuracy (b) of SQMD and FedMD on SC dataset under the asynchronous training condition.}
\label{Fig:case_study}
\vspace{-10pt}
\end{figure} 

\subsection{Asynchronous Setting (\textit{RQ4})}
In real-world healthcare applications, patients in different medical facilities start to train their personal models asynchronously.  Existing on-device heterogeneous learning frameworks simply assume all clients join the learning frameworks at the same training stage. If the clients join the training at different stages/times, newcomers can learn more from other converged models and hence improve their performance faster. However, the knowledge from the newcomers can be ruinous for indigenous clients. In this regard, we conduct a simulation on the SC dataset to reveal the difference between our SQMD and the FedMD framework. We assume there are three medical facilities $M_1$ to $M_3$, and they join the framework one after another. The clients from different facilities have heterogeneous on-device models (i.e., $M_1$ with $10$ ResNet8, $M_2$ with $11$ ResNet20, and $M_3$ with $11$ ResNet50). As time goes by, the test accuracy of the two frameworks is depicted in Fig.~\ref{Fig:case_study}. \hl{The test samples for calculating the overall accuracy come from $\{M_1\}$, $\{M_1, M_2\}$, and $\{M_1, M_2, M_3\}$, and are respectively denoted by Stage $1$, Stage $2$ and Stage $3$.}

\hl{The sudden accuracy drops at the starting points of stage $2$ and stage $3$ in} Fig.~\ref{Fig:case_study_1} \hl{are mainly resulted from the drastic changes in the data. Since the newcomers perform poorly at the beginning while holding a comparable number of test data with other facilities, they significantly bring the average accuracy down. It is noticed that FedMD falls deeper and recovers slower than SQMD, which indicates that the proposed framework is more robust to asynchronous scenarios.} Fig.~\ref{Fig:case_study_2} \hl{urther evidences the impact of the new data samples since the performance of $M_1$ is relatively stable.} It also suggests that well-trained models in SQMD are less affected by cut-in immature members when new facilities affiliate, exhibiting its potential in real-world healthcare applications.

\subsection{\hl{Ablation Study and Covergence Rate Analysis}}

\begin{figure}[t] 
\centering
\vspace{-0pt}
\includegraphics[scale=0.43]{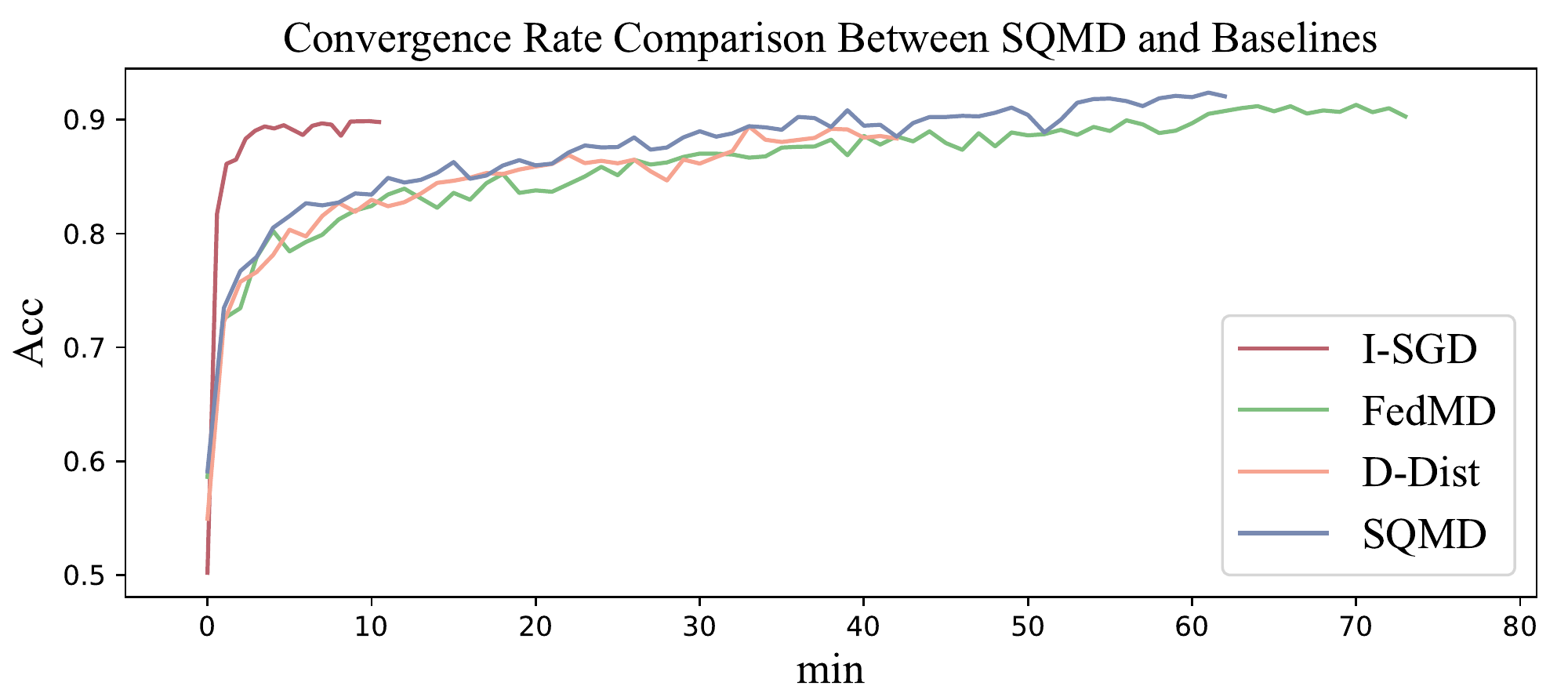}
\vspace{-0pt}
\caption{The test accuracy curve of SQMD and three baselines on PAD dataset.}  
\label{Fig:convergence}
\vspace{-10pt}
\end{figure}

\hl{In this section, we study two variants of SQMD to analyze the performance gains from two kinds of auxiliary information. The first one is SQMD/QF, where we remove the messenger quality filtration mechanism in the server. The second one is SQMD/SF, where the similarity-based neighbor selection procedure is replaced by a random sampling operation (with an equal number of neighbors). The other part of the framework follows the optimal settings in Section}~\ref{sec:Setting}\hl{, and the results on two healthcare datasets are summarized in Table~}\ref{tab:ablation}\hl{. The overall performance has decreased on both tasks after removing one of the components in the framework. This verifies the rationality of introducing two-step messenger filtration in the server. Particularly, we notice that SQMD/QF performs slightly better than SQMD/SF, which implies messenger quality filtration is the major contributor to SQMD's performance gain.

Additionally, the accuracy curves of SQMD and baselines on the PAD dataset are depicted in Fig~}\ref{Fig:convergence}\hl{ to study the convergence rate. The training time of all frameworks is measured by an NVIDIA GeForce RTX 3090 GPU. In this way, we can get rid of the perturbation from unstable communication time costs. I-SGD is the fastest one as expected, but it overfits to its local data and lacks generalization. Notably, compared with the heterogeneous FL baselines FedDM and D-Dist, SQMD obtains a faster convergence rate with a higher final accuracy score, which demonstrates that our messenger-based distillation is advantageous in both effectiveness and efficiency when handling heterogeneous models in FL. Similar observations are drawn on SC and FMNIST datasets, hence we omit plotting their figures due to page limitation.}

\begin{table}[t]
\vspace{-0pt}
\caption{Importance of Key Components in SQMD}
\begin{center}
\begin{tabular}{c|c|c|c|c}
\hline
Method & Metric & SQMD & SQMD/QF & SQMD/SF\\
\hline 
\multirow{3}{*}{SC}
& Acc  & \textbf{0.8465} & 0.8334 & 0.8394 \\
& Pre & \textbf{0.7547} & 0.7283 & 0.7435  \\
& Rec &  \textbf{0.8056} & 0.7802 & 0.7900 \\
\hline
\multirow{3}{*}{PAD} 
& Acc  & \textbf{0.9203} & 0.8846 & 0.9005 \\
& Pre  & \textbf{0.8966} & 0.8912 & 0.8922 \\
& Rec  & \textbf{0.8861} & 0.8465 & 0.8841 \\
\hline
\end{tabular}
\label{tab:ablation}
\end{center}
\vspace{-10pt}
\end{table}

\vspace{-5pt}
\section{Conclusion}\label{sec:conclusion}
In this paper, we design the SQMD, a novel heterogeneous asynchronous collaborative learning framework, to support on-device personalized DNNs training. Boosted by the quality-similarity-based co-distillation communication protocol, our SQMD achieves satisfactory performance compared with existing heterogeneous federated frameworks and decentralized frameworks. Furthermore, our SQMD shows unique superiority when the data is highly sparse, or the training process is asynchronous.

\subsection{Theoretical Implications}
Conventional distributed learning paradigms usually construct static connections guided by external information (e.g., demographics \cite{ye2022personalized}). This will reveal sensitive personal information during knowledge sharing. Meanwhile, high similarity on static external information does not necessarily lead to a performance gain in the collaboration. To this end, we design a novel communication protocol where the connections are constructed based on dynamic model-wise similarity and client model quality. To further protect the privacy of model parameters, we innovatively exploit the auxiliary information coupled in the messengers to calculate the similarity and quality rather than using the local sensitive data or model parameters.

\vspace{-10pt}
\subsection{Practical Implications}
Our SQMD allows each client model to reach its full potential rather than being restricted by the device with the least computing resources in traditional distributed learning paradigms. Meanwhile, data sparsity and asynchronous training environments are two common issues in practical healthcare IoT scenarios. The proposed dynamic-linkage messenger distillation empowers the whole network to resist these two kinds of interference. Experiments on real-life datasets exhibit the immense potential of SQMD in real-life on-device predictive applications.

\section*{Acknowledgment}
This work is partly sponsored by Australian Research Council under the streams of Future Fellowship (FT210100624), Discovery Early Career Researcher Award (DE200101465, DE230101033), Discovery Project (DP190101985), and UQ New Staff Research Start-up Grant (NS-2103).

\ifCLASSOPTIONcaptionsoff
	\newpage
\fi
\bibliographystyle{IEEEtran} 
\bibliography{jbhi}

\end{document}